\documentclass[journal]{IEEEtran}

\usepackage{cite}
\usepackage{amsmath,amssymb,amsfonts}
\usepackage{algorithmic}
\usepackage{textcomp}
\usepackage{xcolor}
\usepackage{soul}
\usepackage{caption}
\usepackage{subcaption}
\usepackage{tikz}
\usepackage{comment}

\usepackage{array,multirow,graphicx}
\newcommand{\STAB}[1]{\begin{tabular}{@{}c@{}}#1\end{tabular}}

\usepackage[ruled,vlined,linesnumbered]{algorithm2e}

\SetKwInput{KwInput}{Input}
\SetKwInput{KwOutput}{Output}
\SetKwRepeat{Do}{do}{while}

\newcounter{todocounter}
\makeatletter\@ifpackageloaded{todonotes}{\newcommand{\todonum}[2][]{\stepcounter{todocounter}\todo[inline]{\thetodocounter: #2\ifx\x#1\x{}\else{ [#1]}\fi}}}{\newcommand{\todonum}[2][]{\par\stepcounter{todocounter}\noindent\fcolorbox{black}{orange}{\begin{minipage}{.98\linewidth}\thetodocounter: #2\ifx\x#1\x{}\else{ [#1]}\fi\end{minipage}}}}\makeatother

\def\BibTeX{{\rm B\kern-.05em{\sc i\kern-.025em b}\kern-.08em
    T\kern-.1667em\lower.7ex\hbox{E}\kern-.125emX}}
\begin{document}

\title{CAPTIVE: \underline{C}onstrained \underline{A}dversarial \underline{P}erturbations to \underline{T}hwart IC Re\underline{v}erse \underline{E}ngineering}

\author{Amir Hosein Afandizadeh Zargari*, Marzieh AshrafiAmiri*, Minjun Seo, Sai Manoj Pudukotai Dinakarrao, Mohammed E. Fouda and Fadi Kurdahi

\thanks{* First two authors contributed equally to this research.}
\thanks{A. Zargari, M. AshrafiAmiri, M. Fouda, M. Seo and F. Kurdahi are with Center for Embedded \& Cyber-physical Systems, University of California-Irvine, Irvine, CA, USA 92697-2625}
\thanks{S. Dinakarrao is with Department of Electrical and Computer Engineering, George Mason University, Fairfax, VA 22030 USA.}
\thanks{Manuscript received xxx, xxx; revised xxxx, xxx.}}

\markboth{
}%
{Zargari \MakeLowercase{\textit{et al.}}: CAPTIVE: Thwarting IC Reverse Engineering. }

\maketitle

\begin{abstract}
Reverse engineering (RE) in Integrated Circuits (IC) is a process in which one will attempt to extract the internals of an IC, extract the circuit structure, and determine the gate-level information of an IC. In general, RE process can be done for validation as well as intellectual property (IP) stealing intentions. In addition, RE also facilitates different illicit activities such as insertion of hardware Trojan, pirate, or counterfeit a design, or develop an attack. In this work, we propose an approach to introduce cognitive perturbations, with the aid of adversarial machine learning, to the IC layout that could prevent the RE process from succeeding. We first construct a layer-by-layer image dataset of 45nm predictive technology. With this dataset, we propose a conventional neural network model called \textit{RecoG-Net} to recognize the logic gates, which is the first step in RE. RecoG-Net is successfully to recognize the gates with more than 99.7\% accuracy. Our thwarting approach utilizes the concept of the adversarial attack generation algorithms to generate perturbation. Unlike traditional adversarial attacks in machine learning, the perturbation generation needs to be highly constrained to meet the fab rules such as Design Rule Checking (DRC) Layout vs. Schematic (LVS) checks. Hence, we propose CAPTIVE as an constrained perturbation generation satisfying the DRC. The experiments shows that the accuracy of reverse engineering using machine learning techniques can decrease from 100\% to approximately 30\% based on the adversary generator.

\end{abstract}


\section{Introduction} \label{intro}








To meet the large operational, maintenance, and development costs, the semiconductor industries are inclining towards a fabless business model i.e., outsourcing the fabrication to offshore foundries. 
Such outsourcing has also lead to other benefits of outsourcing the Integrated Circuit (IC) fabrication and adopting a global supply chain for an 
reduced capital and maintenance costs, minimized design-flow efforts, and time-to-market \cite{James'09,Rakib_Date'21}. 
Despite the achieved benefits, such outsourcing and adoption led to complex verification and fabrication cycle, and supply-chain increased possible hardware threat space, arising in different forms, namely IC piracy, overproduction, hardware Trojan (HT) insertion, and reverse engineering \cite{torrance2011state}.

Reverse engineering (RE) is a process in which one will attempt to extract the internals of an IC, extract the circuit structure, and determine the gate-level information of an IC \cite{quadir2016survey}. 
The RE can lead to adversarial consequences including IP theft, and IP stealing, eventually leading to financial losses \cite{Xiao'16}. 
Among multiple reverse engineering techniques, imaging-based reverse engineering \cite{Randy'09,Ulbert'21} 
is one of the prominent threats which cannot be mitigated, as the fabricated devices are available in the market for consumer and commercial systems. 
The existing imaging-based reverse engineering methods can be divided into destructive and non-destructive approaches \cite{quadir2016survey}. In the destructive method, first scanning electron microscopy (SEM) 
images of different layers of the layout are captured. The obtained IC layout or SEM images are fed to reverse engineering tools such as DeGate \cite{torrance2009state} for reverse engineering and annotation. In contrast, the foundries can also adapt other recently introduced non-destructive imaging techniques such as Ptychographic X-ray laminography, ensuring the reverse-engineered IC functions and not destructed \cite{holler2019three}.

The imaging-based reverse engineering techniques have two steps in common. First of all, 
the gates inside each design should be annotated, then by finding the connectivity between the gates, the whole reverse engineering process is completed, and layout design can be revealed. 
Given the success of machine learning and statistical methods in a wide range of applications \cite{ashrafiamiri2020r2ad,zargari2020newertrack,gnn4tj,zargari2021accurate,gnn4ip,aqajari2021end,aqajari2021pain,yasaei2020iot,ashrafiamiri2018towards} with computer vision and hardware security being no exceptions, we develop a  machine learning model, RecoG-Net is capable of recognizing the type of gate by using the image of one layer of the layout.

Further, as our principal contribution, we propose CAPTIVE, Constrained Adversarial Perturbations to Thwart IC Reverse Engineering. CAPTIVE is capable of making gate recognition, the first step of reverse engineering, impossible.
In the CAPTIVE method, we introduce design-rule checking (DRC) compliant adversarial perturbations which are specially crafted noises and are sometimes invisible to human eyes. The main challenge was to convert the existing adversarial noises \cite{moosavi2016deepfool, papernot2016limitations, andriushchenko2020square} to DRC-compliant objects that can be added to the layout and also be fabricated. An approach is introduced towards reaching these specific DRC-compliant perturbations.
To validate the effectiveness of the CAPTIVE method, DRC-compliant objects are added to the SEM images of layer(s), and the experiments indicate that the RecoG-Net's performance in gate recognition will drastically drop up to 70\% (from near 100\% to 30\%). 

\begin{figure*}[t]
    \centering
    \includegraphics[width=\linewidth]{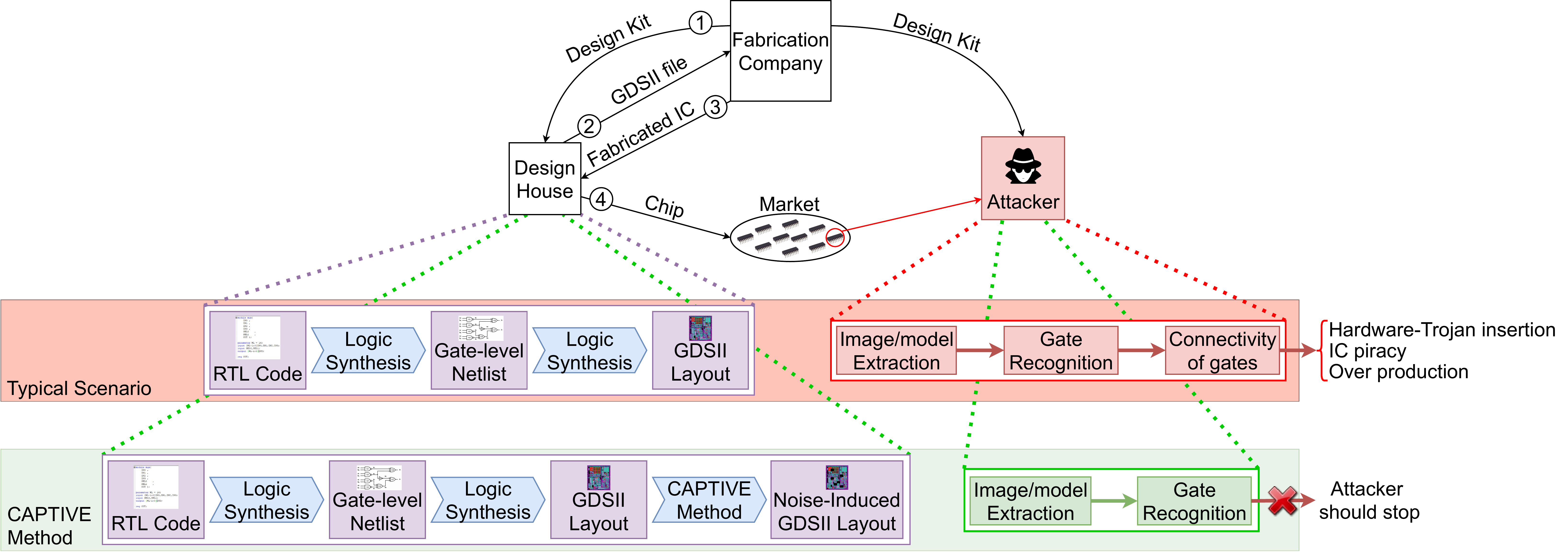}
    \caption{IC reverse engineering attack with and without CAPTIVE.}
    \label{fig:Overview}
\end{figure*}

The reported results present that even with finding the perfect connectivity if an attacker can not recognize gates, the process of reverse engineering will fail. 

The contributions of this work can be outlined in two main points as follows:
\begin{itemize}
    \item We propose RecoG-Net, a convolutional neural network model, to fully recognize the gates from single/multiple layer(s) of SEM or the layout image(s) with approximately 100\% accuracy. 
    
    \item We propose CAPTIVE as a method to add DRC-complaint perturbation to layout images to thwart IC-RE. We perform different experiments to validate the efficiency of CAPTIVE.    
\end{itemize}

The rest of this paper is organized as follows: Section II describes the reverse engineering process and our data generation mechanism. Moreover, the RecoG-Net model has been proposed which is responsible for performing the gate recognition. Section III focuses on the CAPTIVE method to make the first step of reverse engineering, gate recognition, less possible. The evaluation of the proposed CAPTIVE method and some relevant background regarding the existing reverse engineering methods are presented in Section VI. Section V contains the conclusion and the future work.

\section{IC Reverse Engineering Attack}


Nowadays, the fabricated ICs are vulnerable to different malicious behaviors. One of the main possible threats is Reverse Engineering on the fabricated IC. To do the process of reverse engineering on the chip and extract the internal structure inside it, first, gates which are the small components inside the chip, should be determined. Then, by finding the connectivity between different gates, the logic and internal structure of the IC can be determined. 

As our first contribution, we propose the RecoG-Net model which will help us in classifying gates and preventing the first step of IC reverse engineering. To achieve this, first, we need to create a specific dataset of different existing gates.


\subsection{Attack Model}

Integrated Circuit (IC) design refers to a process of assembling a collection of circuit elements like transistors, resistors, and capacitors to perform a specific function. These components are combined to form more complex functions such as logic gates, which are then connected to build more complex segments such as adders and multipliers. This process continues to build on itself, resulting in the availability of increasingly complex circuit building blocks. 

In IC design, the circuit elements are implemented on a silicon substrate using a process called photolithography. The photolithography process creates various geometric shapes on the silicon substrate where the electrical properties of the region defined by that shape are altered. Basic circuit elements are created when these regions are combined and superimposed over each other.

Thus, IC design consists of two distinct processes. First, circuit elements are gathered together in one place to perform some specific pre-defined function. Next, the various geometric shapes that implement those circuit elements must be assembled and interconnected on the silicon substrate. The first process is typically called logic design, and the second process is called physical design. 

Nowadays, several companies accomplish the logic design; but these companies remain fabless due to the high cost of physical design. Therefore, these companies will send the designed circuit to the fabrication company for physical design, which introduces additional challenges, especially reverse engineering, as aforementioned.


The attack model we consider in this work is as follows: 
The attacker has access to a fully functional IC and the goal is to extract the netlist and the internals of the design. In order to achieve this, the attacker performs a destructive de-layering of the design \cite{Randy'11,Lippmann'19,quadir2016survey,Vijaykumar'17}. Based on the de-layered images of the IC, the attacker feeds it to an automated tool such as DeGate \cite{Degate} or a similar surrogate tool to annotate the gates and extracts the high-level information of the IC. A crucial step in the attack process is to identify the individual gates and determine the interconnectivity to further exploit the IP. This process is also outlined in Figure \ref{fig:Overview}.

\subsection{RecoG-Net: Surrogate Reverse Engineering Model}\label{proposedModel}

The primary next step towards reaching our goal is finding a method to determine the type of gate using the image(s) of one or multiple layers. Since each of these images contains a tremendous amount of information about the gate, using machine learning techniques seems to be reasonable. Machine learning techniques are capable of learning the existing pattern in the input data; based on the learned data, they are able to classify or predict a particular outcome for another set of data.

We propose a recognition network, RecoG-Net, a combination of two powerful machine learning techniques, Convolutional Neural Network (CNN) with
fully connected layers. 
Such networks are famous for having an automatic feature selection attitude, which will help reduce the dimension of our input. The dimensionality reduction procedure will both preserve features that are mostly related to the specific characteristics of a gate and eliminate the meaningless ones. This will help the network to learn valuable features thoroughly. 
 
Our experimental dataset consists of 889 images with the size of 258 by 1049 which are converted into black and white images to obtain maximum contrast. The dataset is divided into two parts for the training and testing phases with 700 and 188, respectively. The structure of the model that we have proposed is illustrated in Figure 2. First of all, using a 2D-convolutional layer and 32 filters of size 3$\times$3, the 258$\times$1049 features of the data point are converted to a matrix of 1256$\times$1047$\times$32. This initial layer will learn the basic features of the data. The second 2D-Convolutional layer, which will help learn more complex features, is being used with a similar structure. The pooling layer will slide a filter size of 3$\times$3 across the 254$\times$1045$\times$64 features and replaces it with the maximum value. Therefore it will result in discarding 45\% of the features in the matrix, shaped 84$\times$348$\times$64. To lower the possibility of overfitting, a dropout layer is utilized. The dropout rate is equal to 0.5 in our case, which means that 50\% of the features will be discarded. Another set of convolutional layers and pooling is used for understanding even more complex features. Again, to reduce the likelihood of overfitting, a dropout layer of 0.5 is applied.  A Flatten layer is used to convert the 3-dimensional matrix with the size of 26$\times$114$\times$64 to a 1-dimensional matrix of size 189696. Then, a fully connected layer of size 250 is being followed by the convolutional network. In the end, a dense linear layer with a size equal to 11 will produce the predicted output. More detailed explanations of each layer can be found in table \ref{table:architecture_CNN}. To choose the filters, first, we assign a filter size. Then,  CNN  randomly initializes the filters and trains until it reaches the best filter.  Then we try different filter sizes and do the training by using CNN again. In the end, the filter size with the best result will be chosen as the final filter size.

RecoG-Net is then trained with the training dataset. Then, it's performance is measured based on the testing dataset. The predicted outcome of the network is the type of gate, which is classified into 11 different types. 

\begin{table}
\begin{center}
\caption{RecoG-Net Architecture for gate recognition}
\label{table:architecture_CNN}
\begin{tabular}{c|cc} 
 \hline
 \hline 
  Layer & Structure & Output\\
  \hline
 Conv2d+Relu & 32$\times$3$\times$3 & 256$\times$1047$\times$32 \\
 Conv2d+Relu & 64$\times$3$\times$3 & 254$\times$1045$\times$64 \\
 Max pooling2d & 3$\times$3 & 84$\times$348$\times$64 \\
 Dropout & 0.5 & 84$\times$348$\times$64\\
 Conv2d+Relu & 32$\times$3$\times$3 & 82$\times$346$\times$32 \\
 Conv2d+Relu & 64$\times$3$\times$3 & 80$\times$344$\times$64 \\
 Max pooling2d & 3$\times$3 & 26$\times$114$\times$64 \\
 Dropout & 0.5 & 26$\times$114$\times$64\\
 Flatten &  & 189696\\
 Dense+Relu & 250 & 250\\
 Dense+linear & 11 & 11\\
 \hline
 \hline 
\end{tabular}
\end{center}
\end{table}

\subsection{Training Dataset}


\begin{figure*}[!h]
    \centering
    \begin{subfigure}[b]{0.45\columnwidth}
        \includegraphics[trim=90 35 90 35, clip, width=\linewidth]{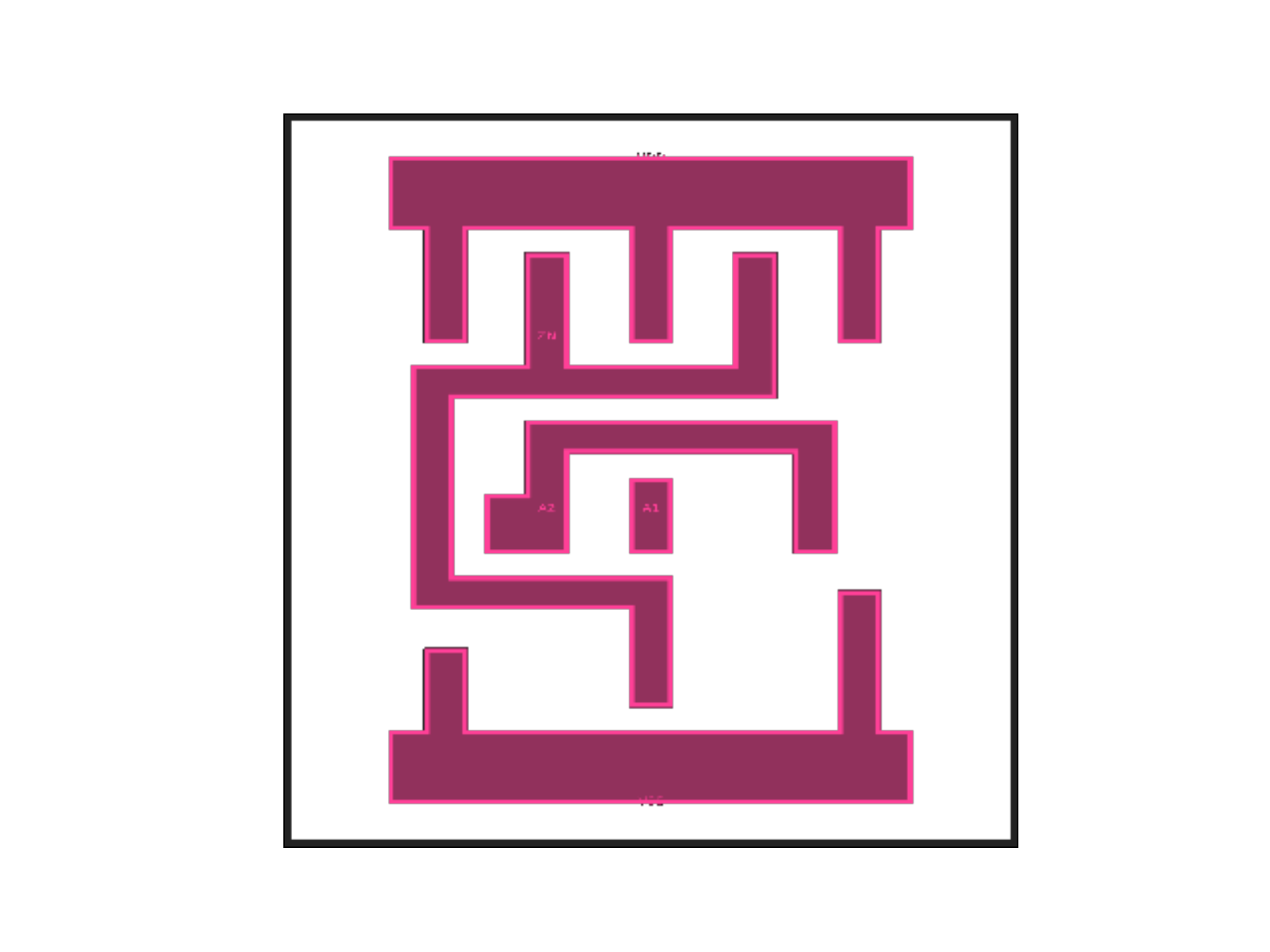}
        \caption{Metal 1 layer}
    \end{subfigure}
    \hfill
    \begin{subfigure}[b]{0.45\columnwidth}
        \includegraphics[trim=90 35 90 35, clip, width=\linewidth]{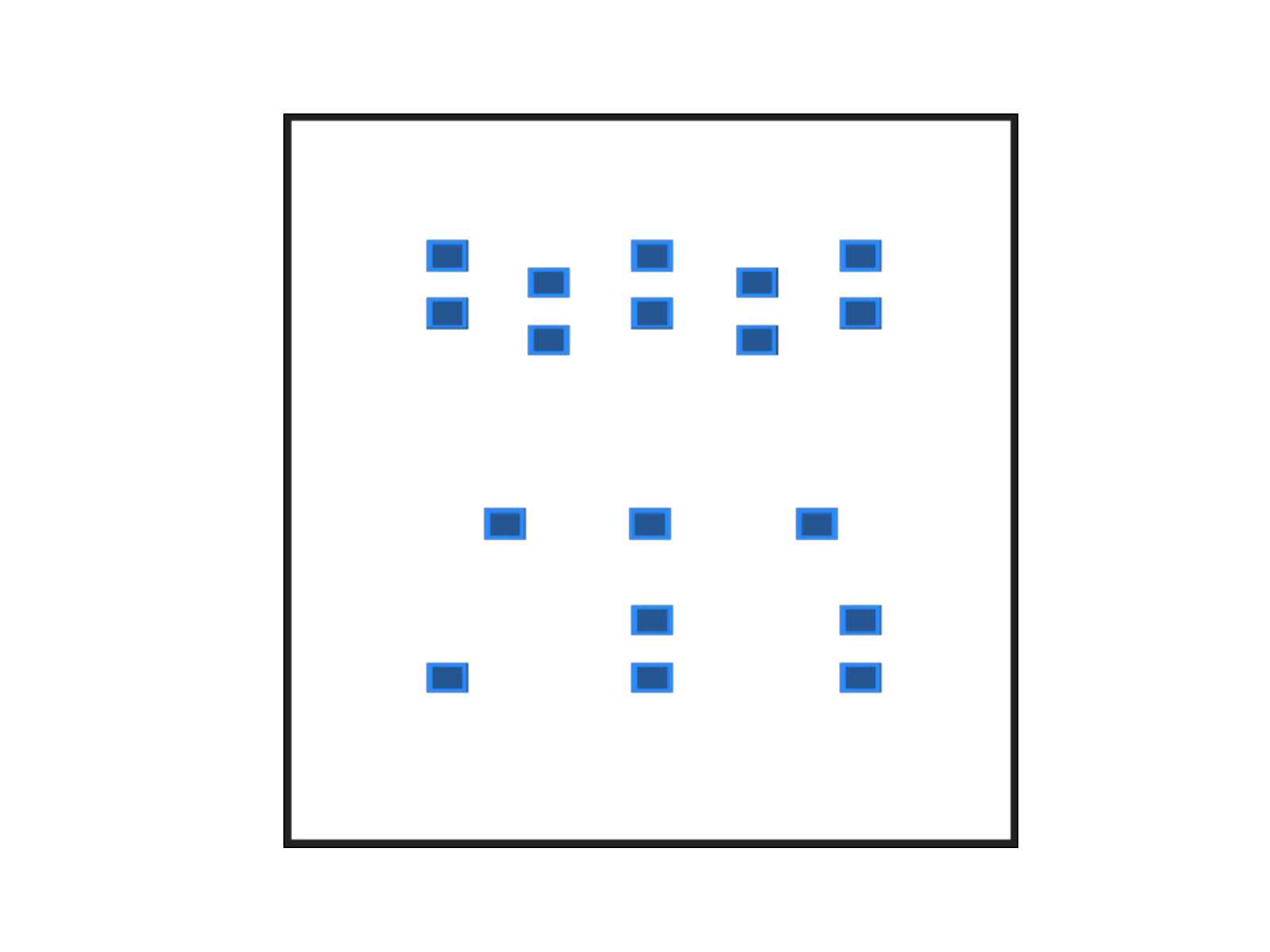}
        \caption{Contact layer}
    \end{subfigure}
    \hfill
    \begin{subfigure}[b]{0.45\columnwidth}
        \includegraphics[trim=90 35 90 35, clip, width=\linewidth]{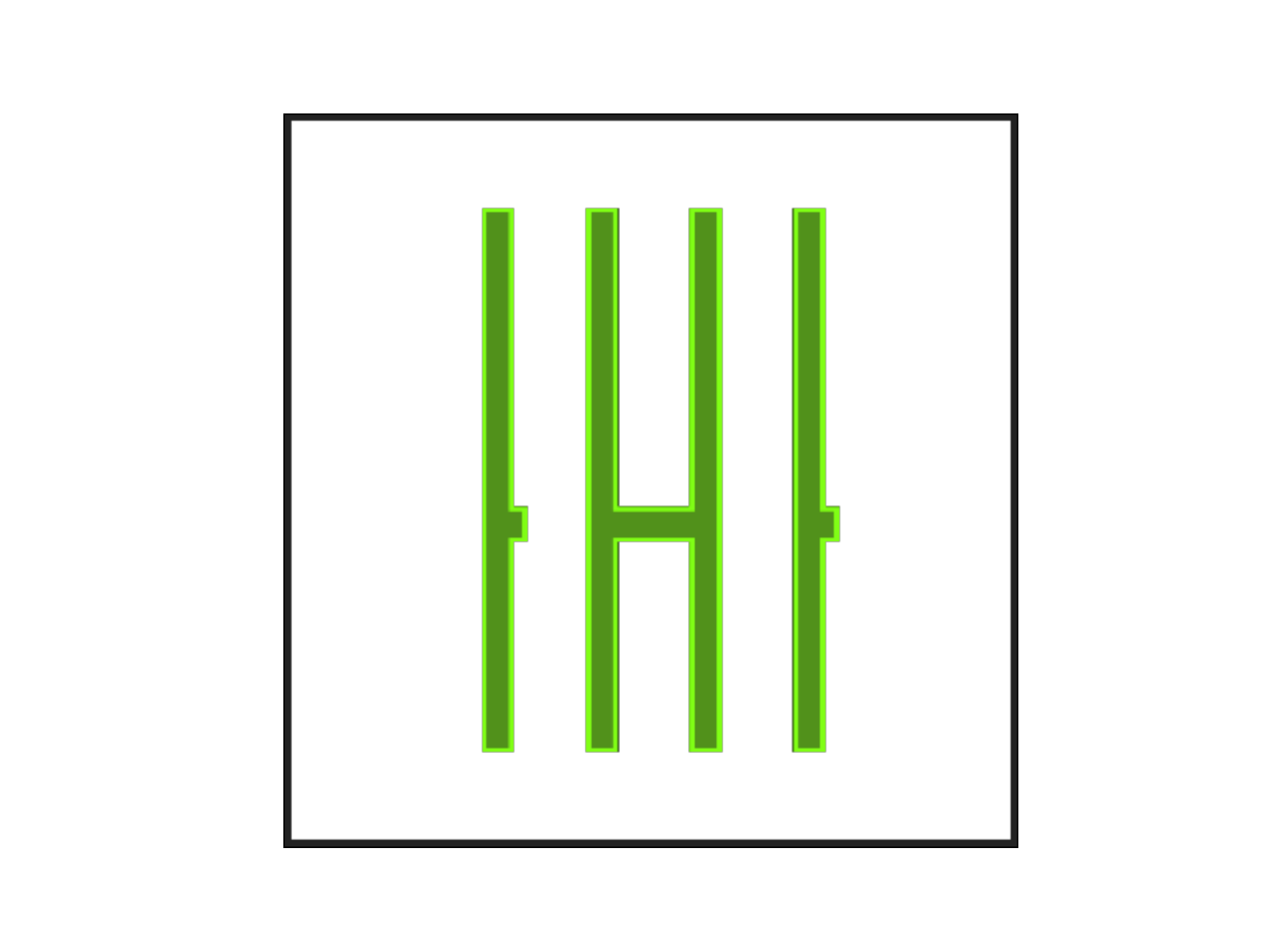}
        \caption{Poly layer}
    \end{subfigure}
        \hfill
    \begin{subfigure}[b]{0.45\columnwidth}
        \includegraphics[trim=90 35 90 35, clip, width=\linewidth]{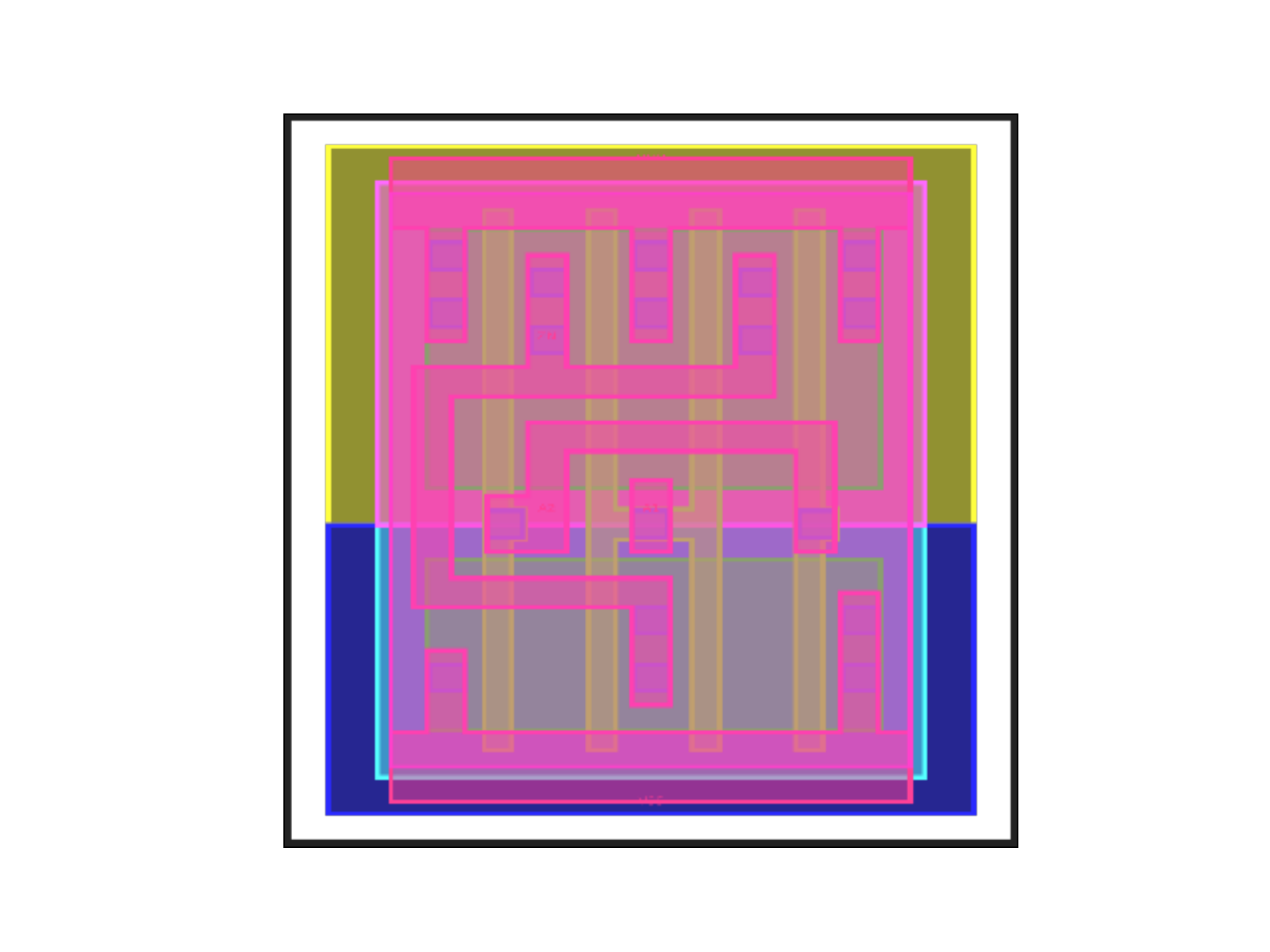}
        \caption{All layers}
    \end{subfigure}
    \caption{An example of different layers of a 2 input NAND gate.}
    \label{fig:NANDexample}
\end{figure*}

The first step towards validating our idea is creating a dataset of all different gates with some specific considered characteristics. The main important feature of the dataset is that it should contain all the essential gates like NAND, NOR, XOR. Moreover, the input size and the number of inputs for each gate should be considered as other parameters.

In addition to the mentioned criteria, having all the different layers existing in each of the gates like the metal layer is needed. 
On the other hand, the dataset that was going to be used in the experiments had to be in image format. After thoroughly checking on the existing datasets, we decided to build up a dataset with all the mentioned features. So, to start our experiments, we used GDSII files of 45nm technology, and extract different layers of each gates. The generated dataset consists of all the layers of all the gates, with different sizing in image format. Figure \ref{fig:NANDexample} shows an example of the constructed dataset of different layers of 2 inputs NAND gate.

\subsection{Fabrication Impact}
As mentioned before, the first contribution of this paper is to find out the type of IC components (gates) using image(s) of one or multiple layers of the gate. To have realistic results, we could not use the generated dataset; Since our created dataset was based on the design layout of the gates, all the edges and corners are entirely straight, accurate, and sharp lines. Figure \ref{fig:NANDexample} is a valid example of this statement.

\begin{figure}[bt]
    \centering
    \includegraphics[width=\linewidth]{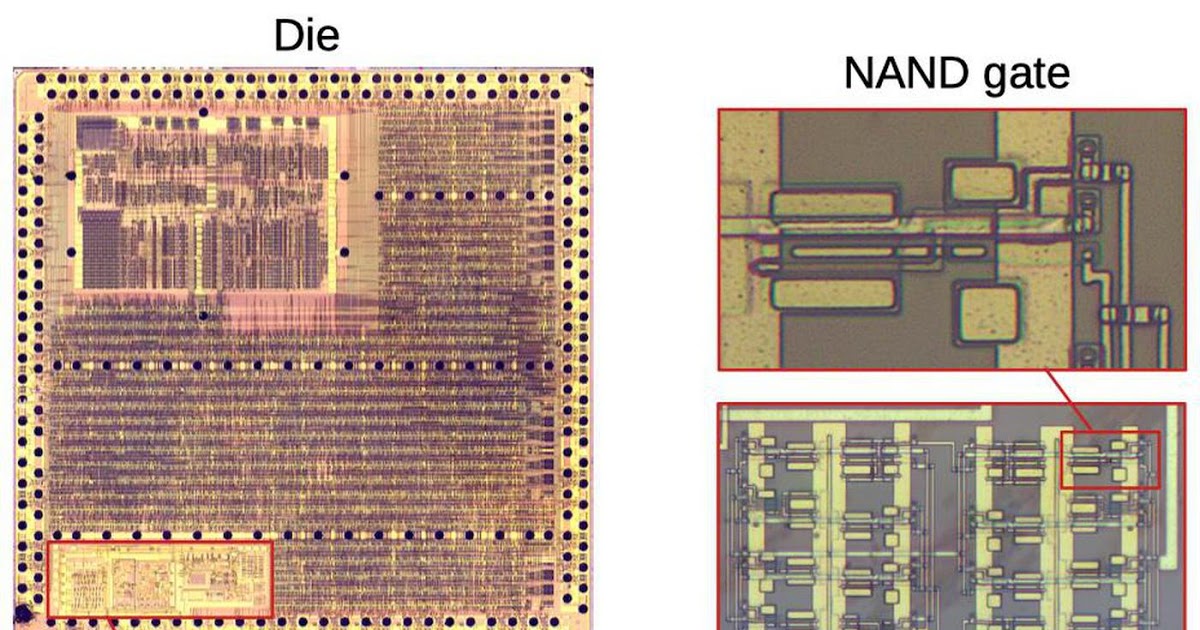}
    \caption{An example of fabricated IC, with a sample of NAND gate inside it.}
    \label{fig:SEMExample}
\end{figure}

\begin{figure*}[t]
    \centering
    \begin{subfigure}[b]{0.19\linewidth}
        \includegraphics[trim=580 110 530 120, clip,width=\linewidth]{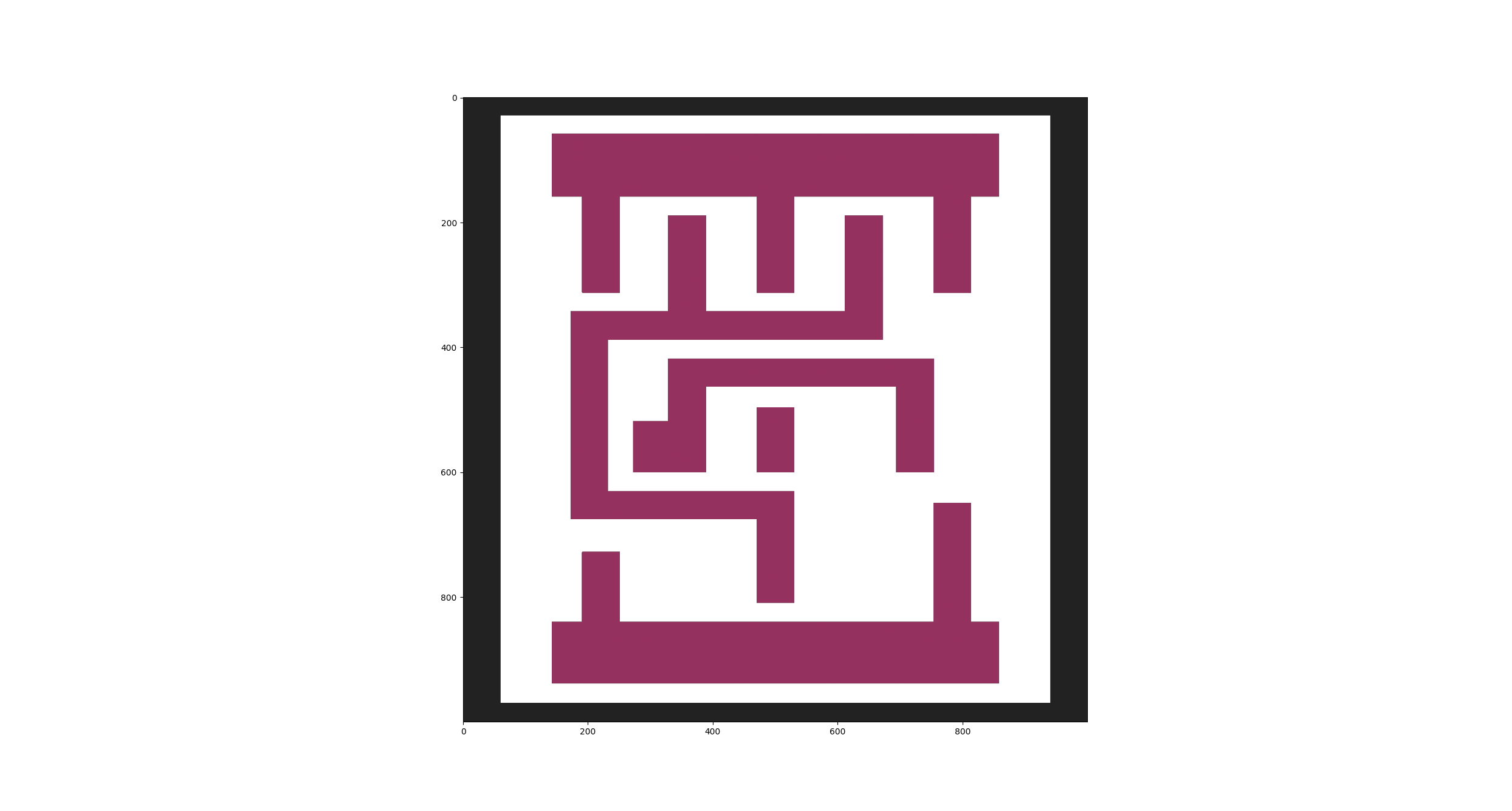}
        \caption{Original}
    \end{subfigure}
    \begin{subfigure}[b]{0.19\linewidth}
        \includegraphics[trim=580 110 530 120, clip,width=\linewidth]{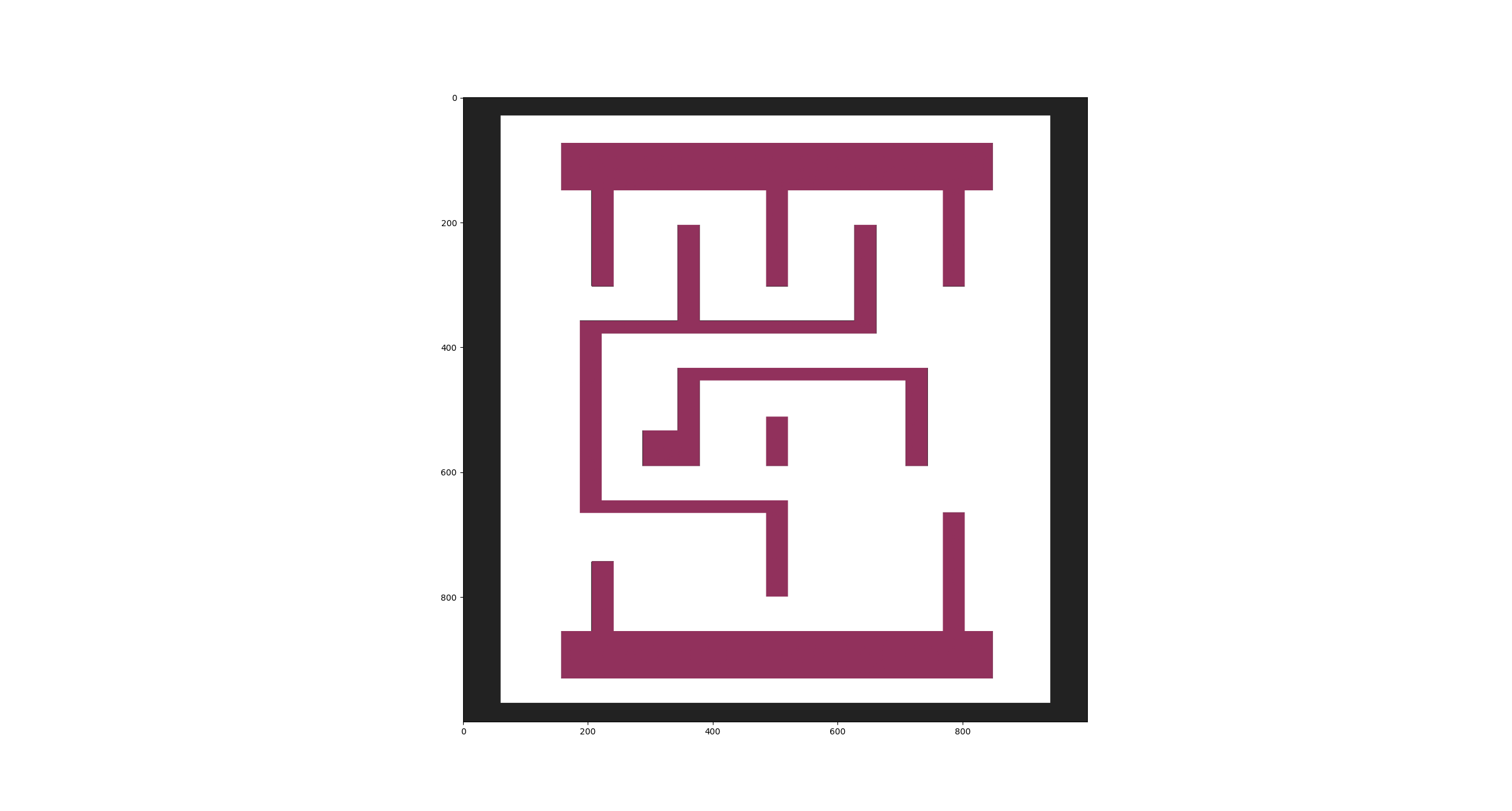}
        \caption{Impact of ``Erosion"}
    \end{subfigure}
    \begin{subfigure}[b]{0.19\linewidth}
        \includegraphics[trim=580 110 530 120, clip,width=\linewidth]{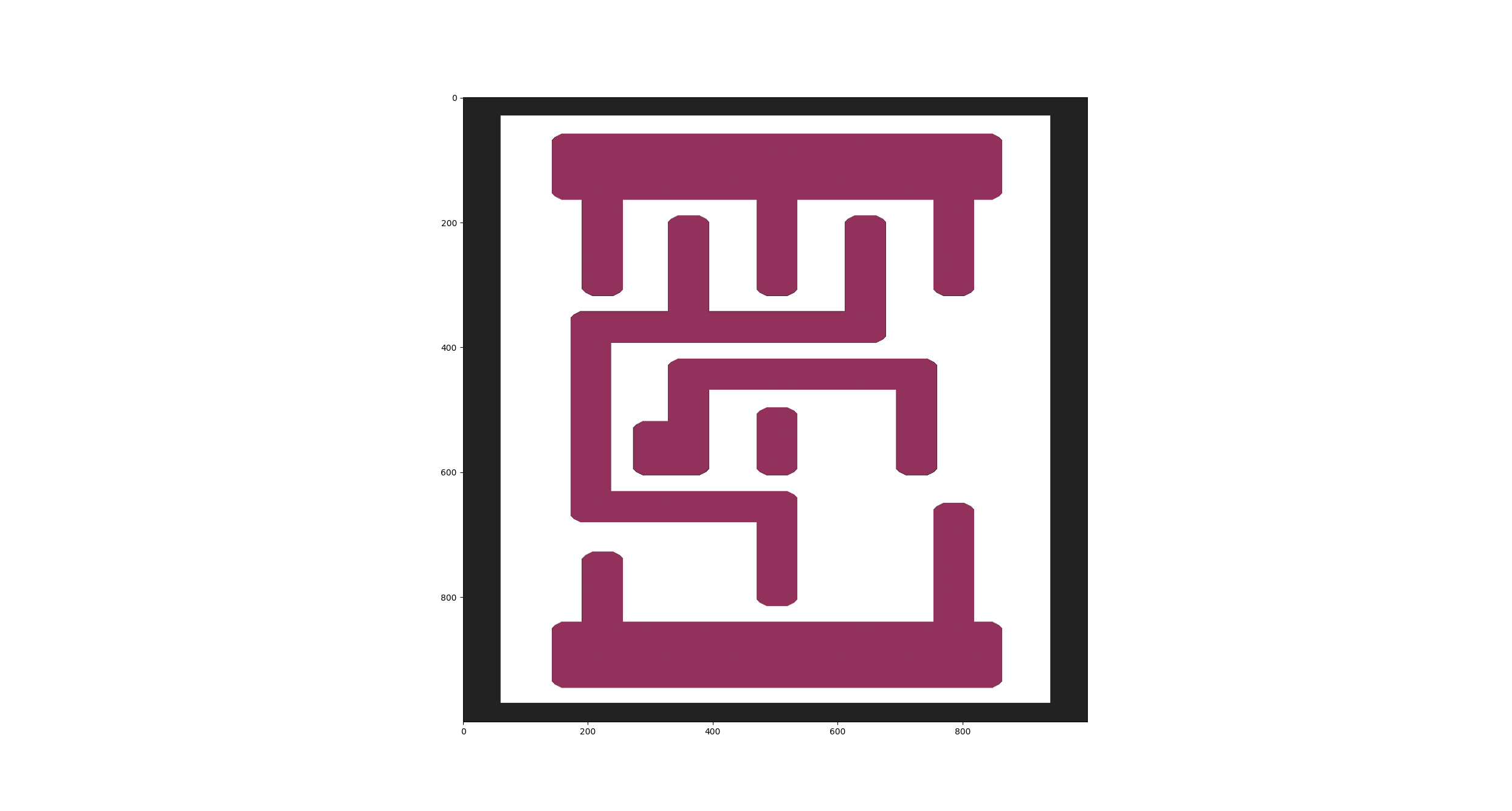}
        \caption{Impact of ``Dilation"}
    \end{subfigure}
    \begin{subfigure}[b]{0.19\linewidth}
        \includegraphics[trim=580 110 530 120, clip,width=\linewidth]{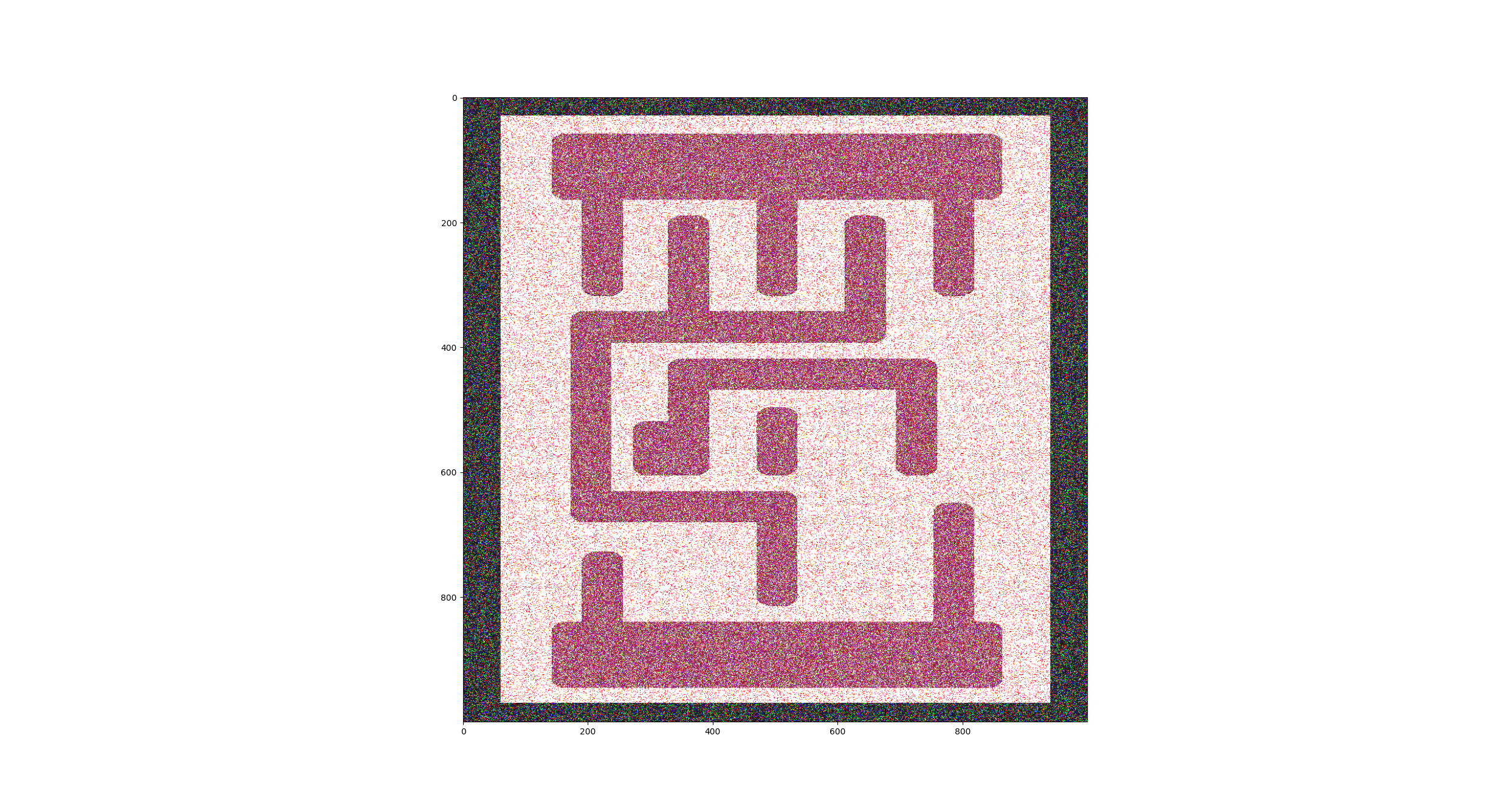}
        \caption{Impact of ``Close"}
    \end{subfigure}
    \begin{subfigure}[b]{0.19\linewidth}
        \includegraphics[trim=580 110 530 120, clip,width=\linewidth]{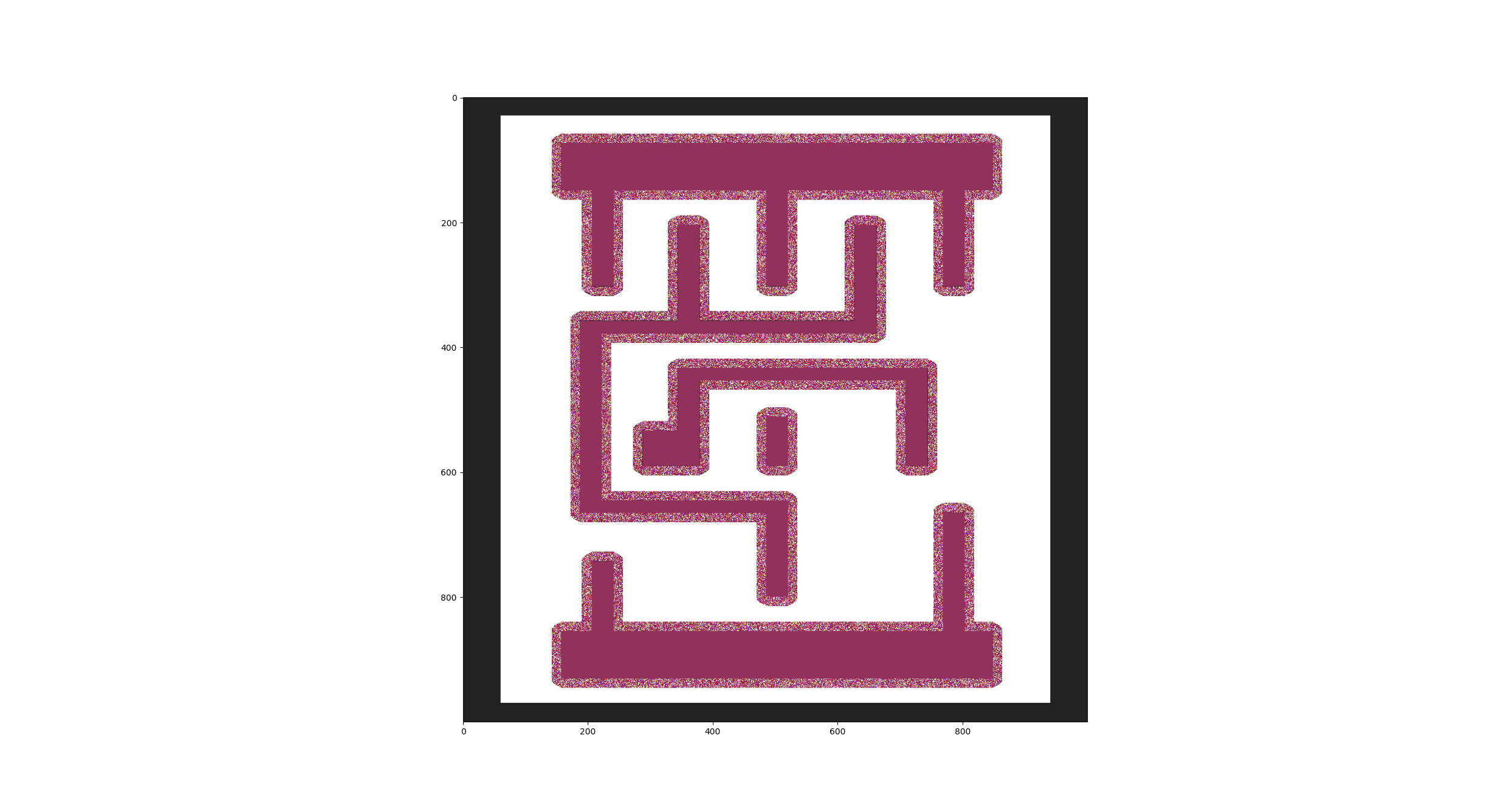}
        \caption{Impact of morphing}
    \end{subfigure}
    \caption{Impact of different morphological transformation operations on a metal layer.}
    \label{fig:morph-operations}
\end{figure*}

Nevertheless, the existing challenge is to distinguish the gates after the physical design and fabrication process. Figure \ref{fig:SEMExample} displays an IC and it's components after fabrication. Therefore, we made some changes in the created dataset to add the effect of fabrication to increase the resemblance of the dataset to an actual fabricated image of IC.

To solve the problem, we applied an image processing technique named morphing transformation. Morphological transformations are collections of some non-linear operations related to the shape or morphology of an image. In other words, these operations do not rely on the numerical values of pixels in the image; just the relative ordering of the pixels is considered. In this technique, two inputs are required, our original image and kernel or the structuring element. The kernel is the input that is responsible for deciding the nature of the ongoing operation. It is positioned at all possible locations in the image, and it is compared with the corresponding neighborhood of pixels. Some operations test whether the element ``fits" within the neighborhood, while others test whether it ``hits" or intersects the neighborhood. Even though different morphological operations exist, no matter what is being used, a morphological operation on an image creates a new image in which the pixel has a non-zero value only if the test is successful at that location in the input image. The two primary operations are Erosion and Dilation, which have been used in this work. 

The Erosion technique erodes away the boundaries of the foreground object. In other words, all the pixels near the boundary of an object will be discarded depending upon the size of the kernel. So the foreground object shrinks. The dilation technique is just the opposite of the erosion technique. So, this technique will increase the thickness or size of the foreground object. In this paper, to mimic the impact of fabrication, the Closing technique of morphology transformation is also used. The closing technique helps to close small holes inside the foreground objects. Figure \ref{fig:morph-operations} indicates the impact of different morphological transformation technique on a metal layer of NAND example.

To create a more realistic dataset, first, we applied the Erosion technique. In parallel, the Dilation technique followed by the Closing process was done. Finally, both of the created images were combined to form a fabricated-resembling dataset.

\section{Adversarial Perturbation Methodology}



We discuss our proposed CAPTIVE technique in this section. CAPTIVE technique eliminates the possibility of reverse engineering a fabricated IC by inducing specially crafted perturbations, named constrained adversarial samples. The perturbed samples include a minimal amount of noise that sometimes can not even be seen by human's naked eyes nor affect the recognition capability by naked human eyes, but can be misclassified by the ML classifiers.

There are lots of different adversarial perturbations. Since the amount of noise added to the image is better to be minimum, we utilize three different perturbations Deepfool, Square-box, and Jacobian-based Saliency Map (JSMA) which are explained in detail in the appendix.

Since the chip images will be fabricated, the noises added to the images must follow the DRC rules. The major problem with the added adversarial noise is that they are not DRC-compliant. In section \ref{attack-generation}, we implemented a method to convert adversarial perturbations to DRC-compliant noises. Later, by using the proposed neural network method in the previous section, we prove that the certainty of the machine learning model in gate recognition would decrease by adding these DRC-compliant noises. Algorithm \ref{alg:validation} dedicates our experimental validation setup. 



\begin{algorithm}
\SetAlgoLined
\KwInput{GDSII file}
\KwOutput{Perturbed GDSII file}
 \Do{High classification accuracy for perturbed data}{
 Initialize perturbation parameter\;
 Generate adversarial perturbation\; 
 Add perturbation Constraints\;
 Test using RecoG-Net\;
 }
 \caption{Experimental validation Setup}
 \label{alg:validation}
\end{algorithm}

\subsection{Perturbation Generation} \label{attack-generation}
Unlike traditional adversarial attacks in machine learning, the perturbation 
generation needs to be highly constrained, such as the perturbation can only be placed in certain parts of SEM/layout to ensure the DRC and LVS checks are not violated and should be more similar to process variations rather than artificially induced.

To force the noise to be DRC-compliant, the first principal rule is to have some pre-defined shapes that can be fabricated. Hence, as the first step, we used a filter with a specific size and moved this filter across the whole image. The filter size is based on the $\lambda$ of the technology size. In each step, we compared the original gate image and the image with induced perturbations. If more than half of the pixels in the filter were perturbed, we considered the whole filter as an induced noise. Since the induced noise with the size of the filter should be fabricated at the end, the filter size shouldn't be too small to cause any problem in fabrication, or too big to be meaningless. 

\begin{figure}[t]
    \centering
    \begin{subfigure}[b]{0.45\linewidth}
        \includegraphics[width=\linewidth]{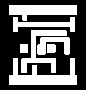}
        \caption{Original image}
    \end{subfigure}
    \hfill
    \begin{subfigure}[b]{0.45\linewidth}
        \includegraphics[width=\linewidth]{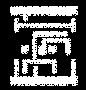}
        \caption{Morphed image}
    \end{subfigure}
    \par\medskip
    \begin{subfigure}[b]{0.47\linewidth}
        \includegraphics[trim=550 110 530 110, clip,width=\linewidth]{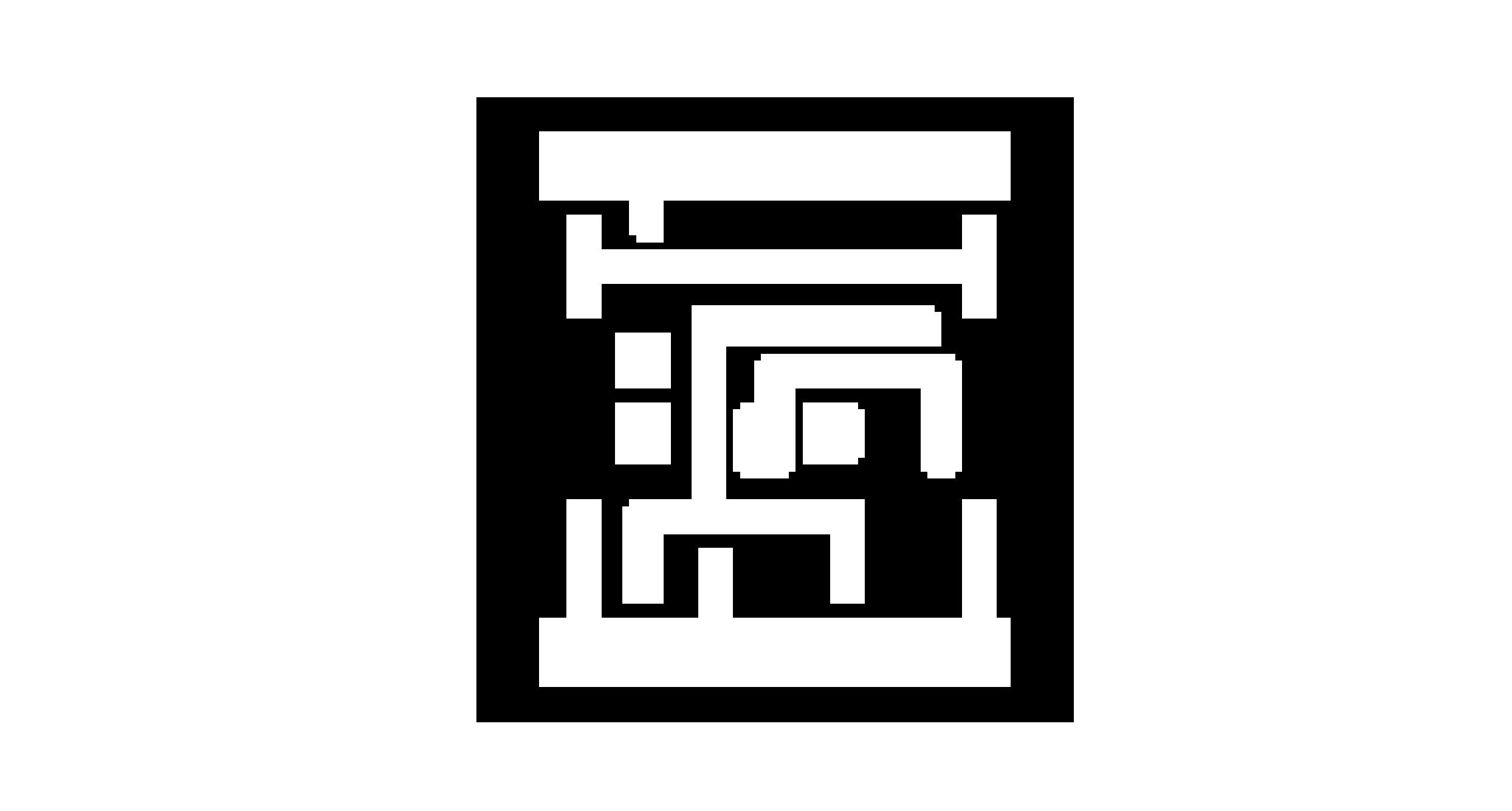}
        \caption{DRC-compliant perturbation added to noise-free image}
        \vspace{3.5mm}
    \end{subfigure}
        \hfill
    \begin{subfigure}[b]{0.45\linewidth}
        \includegraphics[trim=860 410 810 420, clip,width=\linewidth]{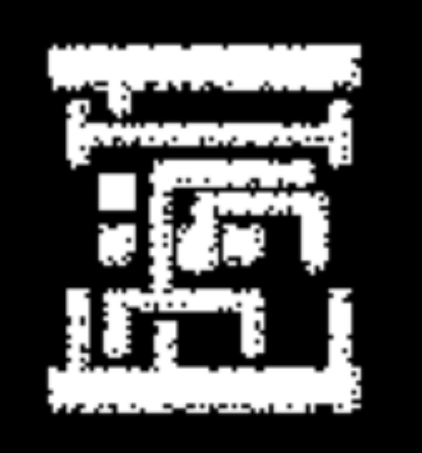}
        \caption{DRC-compliant perturbation added to the morphed image}
    \end{subfigure}

    \caption{The process of converting the original image to an image containing DRC-compliant noise}
    \label{fig:Noise}
\end{figure}

The other essential feature is that all the foreground objects in each layer should have a minimum distance based on the technology size. This minimum distance is $2\times\lambda$. Therefore, the square-shaped induced perturbations can not be added if they have less than $2\times\lambda$ distance from the existing foreground elements in each layer. To achieve this purpose, we defined a forbidden area outside of each object. Figure \ref{fig:Noise} indicates the steps to reach a DRC-compliant noise.

\subsection{Machine Learning Validation}
In section \ref{proposedModel}, we proposed RecoG-Net, a method for gate recognition. In this method, by using image(s) of different layers of gates as an input of DNN and CNN techniques, we identified the type of gate. The idea is first to add the CAPTIVE DRC-compliant noises. Then the chip goes to fabrication and after that RecoG-Net will prove the first step of reverse engineering, gate recognition, is not possible.

By adding the introduced DRC-compliant perturbations in the previous section, we can prove that these added noises can decrease the vulnerability of chip reverse engineering. In other words, this method will cause failure in the first step of chip reverse engineering that is gate recognition. 

To achieve the purpose, RecoG-Net is trained with a noise-free training dataset. The square-shaped DRC-compliant noises are induced in the testing dataset. Then, the performance of the new perturbed dataset can be compared with the performance of the previous noise-free dataset. In the following section, the results of CAPTIVE are presented.

\subsection{DRC and LVS validation}
Design Rule Checker (DRC) checks the layout and verifies if the layout meets all technology-imposed constraints. In comparison to DRC, LVS or Layout Versus Schematic verifies the functionality of the layout. Since there is some induced noise in the GDSII file, it is vital first to pass the DRC rules and then check it with LVS validation tools. 

Since the size of the noise introduces as CAPTIVE is based on the technology size and its position is regarding the constraints with the foreground objects, this noise meets the DRC rules. The next step is to check with some LVS software and verify that CAPTIVE methodology will not change the functionality of the designed gate. 

\section{Results and Discussion}


\begin{table}[]
\centering
\caption{RecoG-Net recognition accuracy results for different layers.}
\label{table:GateRecognition}
\begin{tabular}{|c|c|c|c|c|}
\hline
{Layer name} & {Contact layer} & {Poly layer} & {Metal1 layer} & {All layers}
\\ \hline
{Train Accuracy} & {100\%} & {99\%} & {100\%} & {100\%}\\ \hline
{Test Accuracy}  & {99\%}  & {78\%} & {99\%} & {100\%}\\ \hline
\end{tabular}
\end{table}

\begin{table}[ht]
\caption{Noise-free vs. adversarial vs. DRC-compliant perturbation accuracies with full knowledge of the RecoG-net with 99.8\% and 100\% noise-less accuracies for metal and contact layers, respectively.}
\label{table:AdvDRCAcc}
\centering
\begin{tabular}{|cc|cc|c|}
\hline
\multicolumn{1}{|c|}{Layer}& \begin{tabular}[c]{@{}c@{}}Perturbation \\Method \end{tabular}   & \begin{tabular}[c]{@{}c@{}}Adversarial \\ Accuracy\end{tabular} & \begin{tabular}[c]{@{}c@{}}DRC-compliant\\ Accuracy\end{tabular} & Improvement\\ \hline
\hline
\multirow{3}{*}{\STAB{\rotatebox[origin=c]{90}{Metal}}} &
\multicolumn{1}{|c|}{JSMA}       & 57.5\% & 63\% & 36.8\% \\ \cline{2-5}
&\multicolumn{1}{|c|}{Deepfool}        & 50.1\% & 62.7\% & 37.1\% \\ \cline{2-5}
&\multicolumn{1}{|c|}{Sqaure-box}  &  32.7\% & 38.9\% & \textbf{60.9\%}   \\ \hline
\hline
\multirow{3}{*}{\STAB{\rotatebox[origin=c]{90}{Contact}}} 
&\multicolumn{1}{|c|}{JSMA}           &51\% & 72.4\% & 27.6\% \\ \cline{2-5} 
&\multicolumn{1}{|c|}{Deepfool}       & 62.5\% & 67.7\% & 32.3\% \\ \cline{2-5}
&\multicolumn{1}{|c|}{Sqaure-box}  & 31.9\% & 46\% & \textbf{54\%}   \\ \hline
\end{tabular}
\end{table}

\begin{table}
\begin{center}
\caption{Architectures for Network "A" and "B" in Table
\vspace{-1em}\ref{table:AdvDRCAccBlackboxGeneratedbyRecoG-Net}}
\label{table:architecture_CNN_blackbox}
\begin{tabular}{cc||cc} 
 \hline
 \hline
 \multicolumn{2}{c||}{Network "A"}& \multicolumn{2}{|c}{Network "B"}\\
 \hline 
  Layer & Structure & Layer & Structure\\
  \hline
 Conv2d+Relu & 32$\times$3$\times$3 & Conv2d+Relu & 32$\times$3$\times$3 \\
 Conv2d+Relu & 32$\times$3$\times$3 & Conv2d+Relu & 64$\times$3$\times$3 \\
 Max pooling2d & 3$\times$3 & Conv2d+Relu & 64$\times$3$\times$3 \\
 Dropout & 0.7 & Max pooling2d & 3$\times$3 \\
 Conv2d+Relu & 32$\times$3$\times$3 & Dropout & 0.5 \\
 Conv2d+Relu & 32$\times$3$\times$3 & Conv2d+Relu & 32$\times$3$\times$3 \\
 Max pooling2d & 3$\times$3 & Conv2d+Relu & 64$\times$3$\times$3 \\
 Dropout & 0.7 & Max pooling2d & 3$\times$3 \\
 Dense+Relu & 250 & Dropout & 0.5\\
 Dense+linear & 11 & Dense+Relu & 250\\
  &  & Dense+Relu & 120\\
  &  & Dense+linear & 11\\
 \hline
 \hline 
\end{tabular}
\end{center}
\vspace{-3em}
\end{table}

\begin{table*}[ht]
\caption{Noise-free vs. adversarial vs. DRC-compliant perturbation accuracies with different recognition network. Adversary is generated using RecoG-Net.}
\label{table:AdvDRCAccBlackboxGeneratedbyRecoG-Net}
\centering
\begin{tabular}{|c|c|c|cc|c|c|cc|c|}
\hline
&& \multicolumn{4}{c|}{Network "A"}& \multicolumn{4}{c|}{Network "B"}\\
\cline{3-10}
Layer & \begin{tabular}[c]{@{}c@{}}Perturbation \\ Method \end{tabular}   & \begin{tabular}[c]{@{}c@{}}Noise-free\\ Accuracy \end{tabular}& \begin{tabular}[c]{@{}c@{}}Adversarial \\ Accuracy\end{tabular} & \begin{tabular}[c]{@{}c@{}}DRC-compliant\\ Accuracy\end{tabular} & Improvement & \begin{tabular}[c]{@{}c@{}}Noise-free\\ Accuracy \end{tabular} & \begin{tabular}[c]{@{}c@{}}Adversarial \\ Accuracy\end{tabular} & \begin{tabular}[c]{@{}c@{}}DRC-compliant\\ Accuracy\end{tabular} & Improvement\\ \hline
\hline
\multirow{3}{*}{\STAB{\rotatebox[origin=c]{90}{Metal}}} 
&\multicolumn{1}{|c|}{JSMA} & \multirow{3}{*}{99.9\%} & 64\% & 70.5\% & 29.4\% & \multirow{3}{*}{99.4\%} & 79\% & 81.4\% & 18\% \\\cline{2-2} \cline{4-6} \cline{8-10}
&\multicolumn{1}{|c|}{Deepfool} &  & 52.4\% & 57.1\% & 42.8\% &  & 54\% & 62.4\% & 37\% \\ \cline{2-2} \cline{4-6} \cline{8-10}
&\multicolumn{1}{|c|}{Sqaure-box} &  & 27\% & 34.2\% & \textbf{65.7\%} &  & 39.4\% & 43\% & 56.4\% \\ \hline
\hline
\multirow{3}{*}{\STAB{\rotatebox[origin=c]{90}{Contact}}} &
\multicolumn{1}{|c|}{JSMA} & \multirow{3}{*}{99.4\%} & 51\% & 67\% & 32.4\% & \multirow{3}{*}{100\%} & 67.4\% & 84.5\% & 15.5\% \\ \cline{2-2} \cline{4-6} \cline{8-10}
&\multicolumn{1}{|c|}{Deepfool} &  & 60\% & 67.4\% & 32\% &  & 68\% & 75.4\% & 24.6\% \\ \cline{2-2} \cline{4-6} \cline{8-10}
&\multicolumn{1}{|c|}{Sqaure-box} &  & 24.5\% & 29.6\% & \textbf{69.8\%} &  & 30\% & 36.4\% & 63.6\% \\ \hline
\end{tabular}
\end{table*}

\begin{table}[ht]
\caption{Noise-free vs. adversarial vs. DRC-compliant perturbation accuracies with different recognition network. Adversary is validated using RecoG-Net with 99.7\% test accuracy for metal layer and 99.2\% for contact layer.}
\label{table:AdvDRCAccBlackbox}
\centering
\begin{tabular}{|cc|cc|c|}
\hline
\multicolumn{1}{|c|}{Layer}& \begin{tabular}[c]{@{}c@{}}Perturbation \\ Method \end{tabular}   & \begin{tabular}[c]{@{}c@{}}Adversarial \\ Accuracy\end{tabular} & \begin{tabular}[c]{@{}c@{}}DRC-compliant\\ Accuracy\end{tabular} & Improvement\\ \hline
\hline
\multirow{3}{*}{\STAB{\rotatebox[origin=c]{90}{Metal}}} 
&\multicolumn{1}{|c|}{JSMA}     & 59.2\%  & 78.3\% & 21.4\% \\ \cline{2-5}
&\multicolumn{1}{|c|}{Deepfool}   & 67.4\%  & 76.9\% & 22.8\% \\ \cline{2-5}
&\multicolumn{1}{|c|}{Sqaure-box} & 51\%  & 59.4\% & \textbf{40.3\%} \\ \hline
\hline
\multirow{3}{*}{\STAB{\rotatebox[origin=c]{90}{Contact}}} 
&\multicolumn{1}{|c|}{JSMA}      &  67\%  & 96.2\% & 3\% \\ \cline{2-5}
&\multicolumn{1}{|c|}{Deepfool}       & 44.1\% & 75.5\%      & 23.7\% \\ \cline{2-5}
&\multicolumn{1}{|c|}{Sqaure-box}  & 35.6\%    & 71.8\%   & \textbf{27.4\%} \\ \hline
\end{tabular}
\end{table}

For the first step of the experiments, we created our own dataset consisting of different layers of the existing gates like NAND, NOR, XOR for the 45nm technology. Using the GDSII file for each gate containing the design for all the layers together, we have extracted an image of 258 by 1049 for each layer. To mimic the impact of fabrication on the generated layer images, we applied different morphological transformation techniques.  

\subsection{RecoG-Net Results}\vspace{-1mm}

Since each image consists of just two colors, we converted the images into black and white to get the maximum contrast. After that, by using RecoG-Net, the neural network model introduced in section \ref{proposedModel}, we tried to recognize the gate based on one/multiple layers of gates. We first shuffle the dataset to get more meaningful results; then divide it into training and testing data with nearly 80\% and 20\%, respectively. It is worth mentioning that, due to the shuffling mechanism, we did each experiment 10 times and report the average of all experiments as the final result. So, we trained the model using 80\% of the data and then tested it with other remaining data points. Table \ref{table:GateRecognition} represents the results for using a different layer(s). 

The results proved that RecoG-Net is capable of recognizing gates with 99\% accuracy by using just one layer. This layer can be a metal or a contact layer. Hence, we decided to use these two layers to examine the feasibility of adding perturbations to prevent the first step of reverse engineering. As the results indicate, the image consisting of all layers also has 99\% accuracy. Since this image contains lots of information and is more complex, we decided to continue two parallel experiments using just the metal and the contact layers separately. The simplicity of our dataset will cause the methodology to be more successful.  

\subsection{CAPTIVE Results}

To achieve the primary purpose of this paper, preventing the chip reverse engineering process, we generated different adversarial datasets using different adversarial techniques. First, we applied three different techniques of JSMA, Deepfool, Square, and their combination on our test dataset separately. Then, some square-shaped DRC-compliant noise generating from different adversarial techniques is added to the test dataset. Finally, the performance of RecoG-Net was tested using the newly adversary-generated test data set.

We did a thorough experiment using different possible perturbation parameters for each type of perturbation. Table \ref{table:AdvDRCAcc} presents the best model's performance on different adversarial techniques with different amounts of perturbations that we used. For JSMA, $\gamma$ and $\theta$ are set to 1 and 10, respectively. For deepfool, the amount of perturbation is considered 0.5. For square-box attack, the best possible parameters are set to $norm=0, max\_iter=500, \epsilon=0.1, p\_init=0.7, nb\_restart=3, batch\_size=64$. These parameters are similar for both metal and contact layers.

The results in Table \ref{table:AdvDRCAcc} shows that the most effective perturbation method for both metal and contact layers is square-box where for the metal layer the perturbation accuracy dropped to 32.7\% and the DRC-compliant accuracy is 38.9\%. Also, for the contact layer, the perturbation accuracy and DRC-compliant accuracy are equal to 31.9\% and 46\%. In addition, the square-box is found to perturb the smallest number of pixels among the other techniques.
This makes the square-box perturbation is the best candidate for further study to improve the results.   


The aforementioned results are performed assuming that the designer has full knowledge of the attacker model. In other words, the same network (i.e, RecoG-Net) classifying the gates is used to generate the perturbations. Thus, Table \ref{table:AdvDRCAcc} shows the best achievable results. This scenario is rare to happen since usually the attacker and the designer have no relation. Hence, we consider two other scenarios where the attacker has no knowledge of the perturbation network. 

In the first scenario, the DRC-compliant perturbations are generated using the RecoG-Net model, and the generated images are validated using some different neural network model. All the parameters used for generating adversaries are the same as before. Table \ref{table:AdvDRCAccBlackboxGeneratedbyRecoG-Net} shows the results for two of the networks among many distinct networks that we experimented on. The structure for both networks are in table \ref{table:architecture_CNN_blackbox}. The reason behind choosing these two specific networks is based on their performance on gate recognition when adding DRC-compliant perturbations. Based on the results expressed in table \ref{table:AdvDRCAccBlackboxGeneratedbyRecoG-Net}, network "A" has the lowest accuracy in gate recognition compared to network "B", with the highest accuracy in gate recognition. The results in table \ref{table:AdvDRCAccBlackboxGeneratedbyRecoG-Net} illustrates that the most reduction will happen if we apply square-box adversarial perturbation. The possible improvement can be as high as 65.8\% or as low as 18\% for metal layer (respectively 70.4\% and 15.5\% for contact layer) regarding the network being used. Even with the lowest improvement in this range, the accuracy of gate recognition is around 80\% which means that gate recognition can not be done completely.

For the second scenario, we have designed several CNNs similar to RecoG-Net and generate the DRC-compliant perturbations using these networks. Afterward, the images were validated using RecoG-Net. The results of one of them are depicted in Table \ref{table:AdvDRCAccBlackbox}. All the parameters used for each of the perturbations are the same as previously mentioned. Table \ref{table:AdvDRCAccBlackbox} shows that the most reduction in the accuracy happens if the square-box adversarial perturbation is applied, followed by the CAPTIVE method. Clearly, when the perturbation network and attacker network as mismatched, the improvement is dropped by around 40\%.

\subsection{Related Work}

In this section, we briefly overview three main tools for IC reverse engineering and discuss the capabilities of each tool.

One of the first successful trials for IC reverse engineering is Degate tool \cite{torrance2009state}. Degate is a semi-automated tool to assist in reverse-engineering the digital logic in ICs. Degate tool performs three main steps; template matching to recognize gates where it matches standard cells on the imagery given by graphical template then Via matching followed by wire matching. Then, the SPICE netlist has to be constructed manually. 

\textit{Pix2Net} is another powerful tool to extract the schematic and logic gates from SEM images  from \textit{MicroNet Solutions Inc}\cite{pix2netmanual}. The primer version comes with full circuit extraction, VHDL, or SPICE netlisting with a full schematics library. During the schematics extraction, the Pix2Net performs many steps including layer alignment, stitching, auto cell identification, placement, and electrical error checking, and others.

Another commercial and powerful tool is \textit{Circuit Vision} from \textit{TechInsights}. This tool works from IC level reverse engineering going up to the system level and functionality. It also supports different types of circuits and chips including analog, digital, MEMS, and others. The tool was tested on many commercial memory chips and major companies including Samsung, SK Hynix, Micron, and Intel. 

\section{Conclusion and Future Work}

This paper demonstrates an experiment about the feasibility of adding DRC-compliant noise to the GDSII file to prevent the first step towards chip reverse engineering, gate recognition. First, we created the dataset containing the images of different layers of different gates using 45nm technology. Then, we classify them into 11 categories and developed RecoG-Net, a neural network model to do the gate recognition based on the created dataset. As our last and foremost contribution, we have generated DRC-compliant perturbations based on some of the adversarial perturbations. These perturbations are added to the images of different layers. Our experiments report that by adding these DRC-compliant to the images of the metal and contact layers, the performance of gate recognition will drastically drop, which will cause unsuccessful gate recognition as the first step of reverse engineering. 

To further improve our methodology, a surrogate gradient perturbation has to be formulated with constraints on the physical design rules. Square-box problem formulation would be a good candidate to start with since it shows the best perturbation accuracy among the investigated methods. Secondly, a combination of different adversarial perturbations can be a good candidate for enhancing CAPTIVE methodology.

\appendix
The following are summary of the used perturbations methods in CAPTIVE.

\vspace{-3mm}
\subsection{Jacobian-based Saliency Map Attack (JSMA)}
In contrast to applying noise to every single feature of the input data, \cite{papernot2016limitations} proposes an iterative technique 
to add the perturbation, where the forward derivative of DNN is exploited for adding the perturbations. 

Consider a neural network $F$ with input $x$.  
If the corresponding output is class $j$, we represent the 
model as $F_j(x)$. 
The main principle of this work is: to provide target $t$ 
as the output, the probability for $F_t(X)$ must be increased, and simultaneously, the probabilities of $F_j(X)$ for all the other classes i.e., $j \neq t$ have to be decreased, until $t=arg\ max_{j} F_j(X)$ is achieved. This is accomplished by exploiting the saliency map, as defined below
\begin{equation}
S(X,t)[i]= \begin{cases}
0, \text{if} \frac{\partial F_t(X)}{\partial X_i} < 0 \  \text{or} \sum_{j\neq t} \frac{\partial F_j(X)}{\partial X_i} > 0\\ 
(\frac{\partial F_t(X)}{\partial X_i})|\sum_{j\neq t} \frac{\partial F_j(X)}{\partial X_i}|, \text{otherwise}
\end{cases}
\end{equation}
For an input feature $i$ starting with the normal input $x$, we determine the 
pair of features $\{i,j\}$ that maximizes $S(X, t)[i] + S(X, t)[j]$ and perturb 
each of the features by a constant offset $\epsilon$. 
This process is repeated iteratively 
until the target misclassification is achieved.

\vspace{-4mm}\subsection{DeepFool }
DeepFool (DF) is an untargeted adversarial attack optimized for $L_2$ norm, introduced in \cite{moosavi2016deepfool}. 
DF is an efficient adversarial attack that is capable of producing adversarial samples that highly resemble the original inputs as compared to the aforementioned adversarial samples, especially FGSM and BIM attacks. The principle of the Deepfool attack is to assume neural networks as completely linear with a hyperplane separating each class from another. 
Based on this assumption, an optimal solution to this simplified problem is derived from constructing adversarial samples. 
As the neural networks are non-linear in reality, 
the same process is repeated considering the non-linearity into the model. This process is repeated multiple times for creating adversaries. This process is terminated when an adversarial sample is found, i.e., misclassification happens. 

\vspace{-4mm}\subsection{Square-box Attack}
In contrast to the other adversarial attacks, which primarily rely on the gradients to insert the perturbations, 
Square-box attack \cite{andriushchenko2020square} is non-gradient attacks, which performs a randomized search and inserts square-shaped 
perturbations, and in each iteration, the perturbation is situated approximately at the boundary of the images. 
Square attack requires fewer queries for inserting the square pixels compared to other attacks due to the random search and sampling distribution information. 
\vspace{-6mm}

\section*{Acknowledgement}
The authors would like to thank Monisha Loganathan for helping to construct the dataset from the technology kit.

\bibliographystyle{ieeetr}
\bibliography{literature}

\end{document}